\documentclass[aps,prb,twocolumn,showpacs,showkeys,groupedaddress,floatfix,amsmath,amssymb]{revtex4}
\usepackage{graphicx}

\begin{document}

\title{Dynamics of the superconducting condensate in the presence of a magnetic field.
 Channelling of vortices in superconducting strips at high currents}

\author{D. Vodolazov}
\email{denis@uia.ua.ac.be}
\author{B.J. Baelus}
\author{F.M. Peeters}
\email{peeters@uia.ua.ac.be}
\affiliation{Departement Natuurkunde,
Universiteit Antwerpen (Campus Drie Eiken),
\\ B-2610 Antwerpen, Belgium}

\begin{abstract}
On the basis of the time-dependent Ginzburg-Landau equation we
studied the dynamics of the superconducting condensate in a wide
two-dimensional sample in the presence of a perpendicular magnetic
field and applied current. We could identify two critical
currents: the current at which the pure superconducting state
becomes unstable ($J_{c2}$ \cite{self1}) and the current at which
the system transits from the resistive state to the
superconducting state ($J_{c1}<J_{c2}$). The current $J_{c2}$
decreases monotonically with external magnetic field, while
$J_{c1}$ exhibits a maximum at $H^*$. For sufficient large
magnetic fields the hysteresis disappears and $J_{c1}=J_{c2}=J_c$.
In this high magnetic field region and for currents close to $J_c$
the voltage appears as a result of the motion of separate
vortices. With increasing current the moving vortices form
'channels' with suppressed order parameter along which the
vortices can move very fast. This leads to a sharp increase of the
voltage. These 'channels' resemble in some respect the phase slip
lines which occur at zero magnetic field. \vspace{1pc}
\end{abstract}

\maketitle

It is well-known that the resistive state in superconducting wires
or stripes with diameter/width less or comparable to the coherence
length $\xi$ is realized through the appearance of phase slip
centers \cite{Ivlev,Tinkham1,Tidecks}. The phenomenological theory
of phase slip centers (PSC) was first proposed in Ref.
\cite{Skocpol}. According to this theory a PSC is a region with
size of order $\xi$ where the order parameter is strongly
suppressed. The normal current density produced by the oscillation
of the order parameter in the phase slip center decays on a larger
distance scale $\Lambda_Q\gg \xi$ \cite{Skocpol}.

Current-voltage characteristics of such a system is usually
irreversible \cite{Ivlev,Tidecks}. It is possible to distinguish
two critical current densities: $j_{c2}$ - current density at
which the superconducting state becomes unstable and
$j_{c1}<j_{c2}$ - current density below which the phase slip
solution does not exist in the system \cite{Ivlev,Tidecks}. In
recent work \cite{Michotte} it was claimed that the current
$j_{c1}$ is defined by the competition between two relaxation
times: the relaxation time of the absolute value of the order
parameter $\tau_{|\psi|}$ and the relaxation time of the phase of
the order parameter $\tau_{\phi}$. The phase slip solution does
exist when roughly $\tau_{|\psi|}>\tau_{\phi}$.

Because an applied magnetic field suppresses the order parameter
and leads to the appearance of screening currents in the sample it
is naturally to expect that it affects the phase slip process in
the superconductor. For example in Refs. \cite{Kadin,Michotte} it
was shown that a parallel (to the direction of the injected
current) magnetic field modifies the critical currents $j_{c1}$,
$j_{c2}$ \cite{Michotte} and the stair structure of the
current-voltage characteristic \cite{Kadin}.
\begin{figure}[hbtp]
\includegraphics[width=0.45\textwidth]{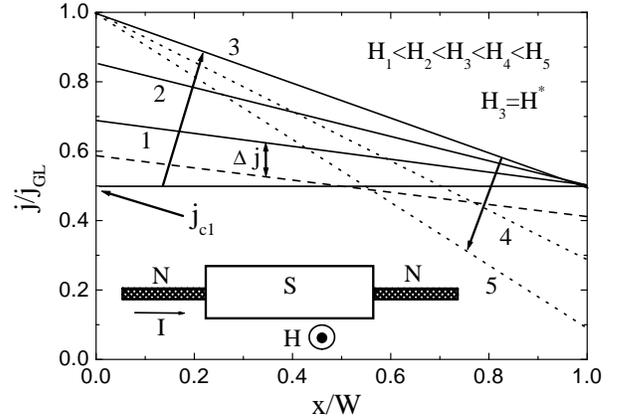}
\caption{Schematic current density distribution in a
superconducting stripe in a perpendicular magnetic field at the
critical currents $I_{c1}(H)$ (solid lines) and $I_{c2}(H)$
(dotted lines). In the insert a schematic view of the considered
set up is shown.}
\end{figure}

If we apply a perpendicular magnetic field the situation becomes
more complicated. In this case the screening currents induced by
the magnetic field decreases the current density j on one side of
the superconductor and increases j on the other side of the sample
(see Fig. 1) and the current density in some part of the stripe
becomes smaller than $j_{c1}$ (if total current is equal to
$J=j_{c1}W$). In accordance with Ref. \cite{Michotte}, in that
part of the sample the phase slip process cannot be realized. We
should increase the applied current in order to satisfy the
condition $j>j_{c1}$ in any point of the line connecting the two
opposite sides of the strip. Further increasing H the slope of
j(x) increases (at current $J=J_{c1}(H)$) and at the moment when
the current density on the left side of the sample reaches
$j_{c2}$ (see Fig. 1) both critical currents becomes equal to each
other $J_{c2}=J_{c1}$ at field $H=H^*\sim (j_{c2}-j_{c1})/W$.
\begin{figure}[hbtp]
\includegraphics[width=0.45\textwidth]{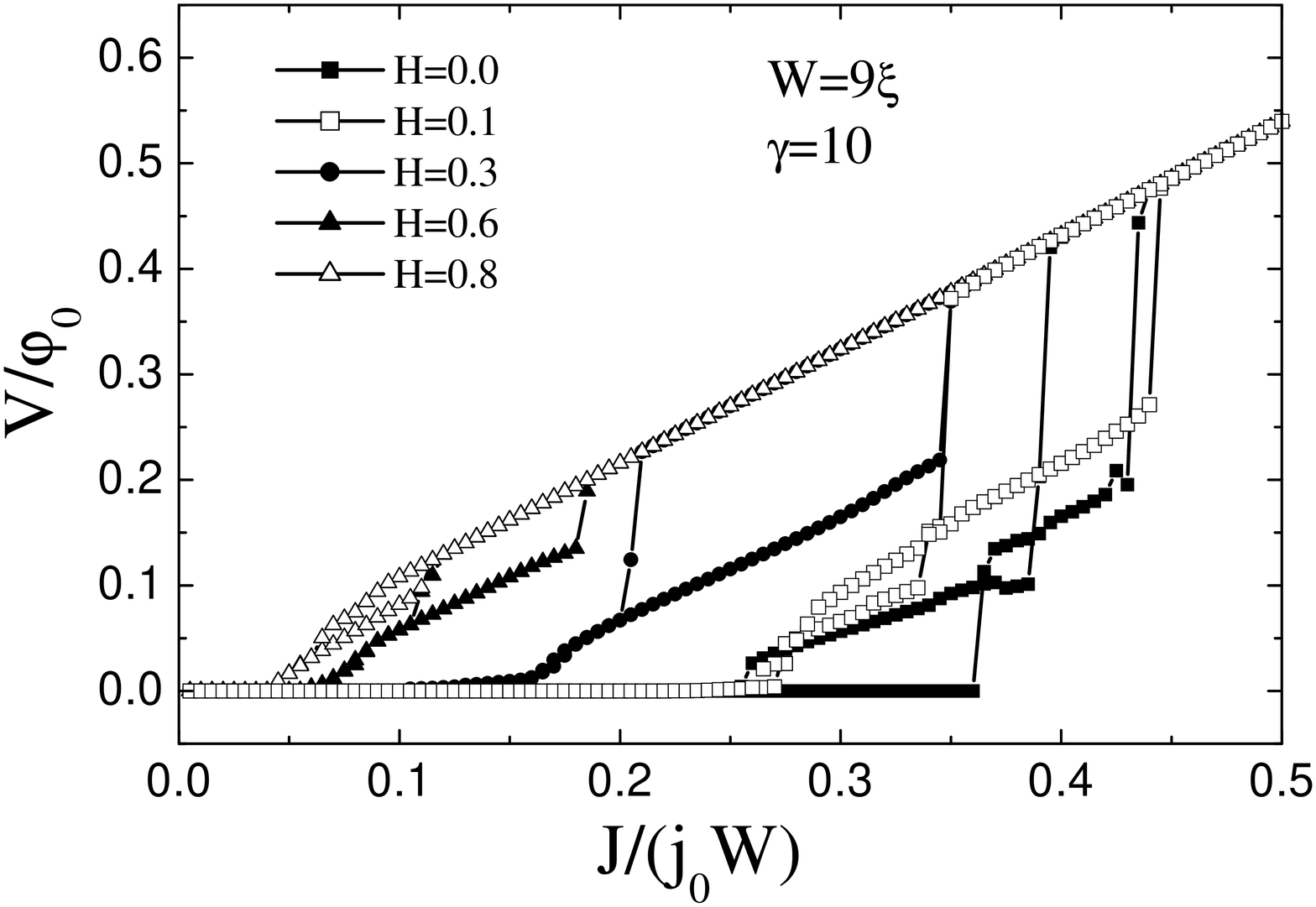}
\caption{Current-voltage characteristics of a superconducting
strip in a perpendicular magnetic field for sweep up and sweep
down regimes.}
\end{figure}

To check the above predictions we studied the current-voltage
characteristics of quasi-two-dimensional superconductors using the
generalized time-dependent Ginzburg-Landau (TDGL) equation. The
latter was first written down in the work of Ref. \cite{Kramer}
\begin{eqnarray}
\frac{u}{\sqrt{1+\gamma^2|\psi|^2}} \left(\frac {\partial
}{\partial t} +i\varphi
+\frac{\gamma^2}{2}\frac{\partial|\psi|^2}{\partial t}
\right)\psi= \nonumber \\
=(\nabla - {\rm i} {\bf A})^2 \psi +(1-|\psi|^2)\psi.
\end{eqnarray}
and should be supplemented with the equation for the electrostatic
potential
\begin{eqnarray} \Delta \varphi & = & 
{\rm div}\left({\rm Im}(\psi^*(\nabla-{\rm i}{\bf A})\psi)\right),
\end{eqnarray}
which is nothing else than the condition for the conservation of
the total current in the wire, i.e. ${\rm div} {\bf j}=0$. In Eqs.
(1,2) all the physical quantities (order parameter
$\psi=|\psi|e^{i\phi}$, vector potential ${\bf A}$ and
electrostatic potential $\varphi$) are measured in dimensionless
units: the vector potential ${\bf A}$ and momentum of
superconducting condensate ${\bf p}=\nabla \phi -{\bf A}$ is
scaled by $\Phi_0/(2\pi\xi)$ (where $\Phi_0$ is the quantum of
magnetic flux), the order parameter is in units of $\Delta_0$ and
the coordinates are in units of the coherence length $\xi(T)$. In
these units the magnetic field is scaled by $H_{c2}$ and the
current density by $j_0=c\Phi_0/8\pi^2\Lambda^2\xi$. Time is in
units of the Ginzburg-Landau relaxation time
$\tau_{GL}=4\pi\sigma_n \lambda^2/c^2=2T\hbar/\pi\Delta_0^2$, and
the electrostatic potential ($\varphi$) is in units of
$\varphi_0=c\Phi_0/8 \pi^2 \xi \lambda \sigma_n=\hbar/2e\tau_{GL}$
($\sigma_n $ is the normal-state conductivity). The parameter $u$
is about $5.79$ according to Ref. \cite{Kramer}. We also put ${\bf
A}=(0,Hx,0)$ in Eq. (1,2) because we considered the case when the
effect of the current-induced magnetic field is negligible.
\begin{figure}[hbtp]
\includegraphics[width=0.45\textwidth]{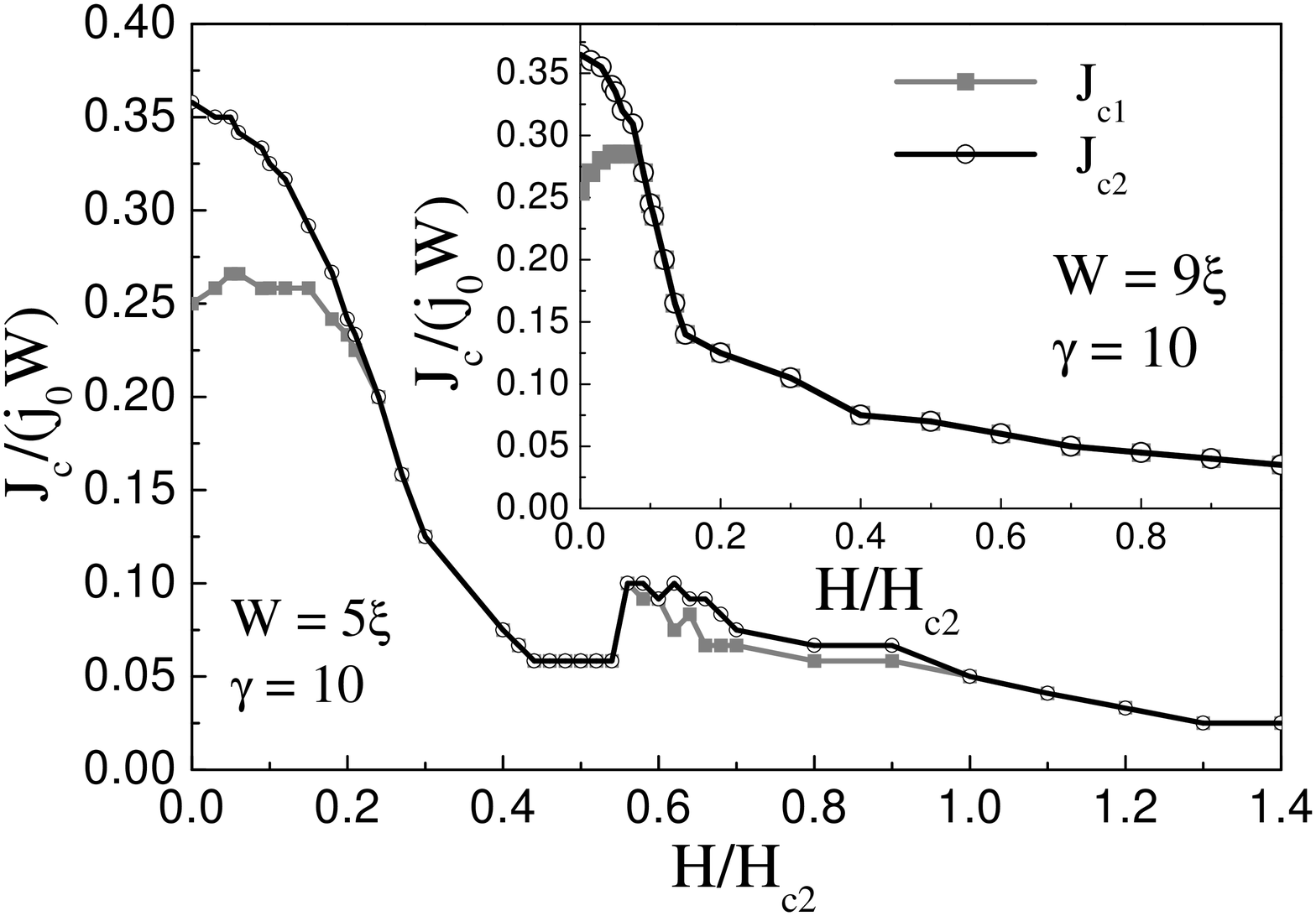}
\caption{Dependence of the critical currents $j_{c1}$ and $j_{c2}$
on the applied perpendicular magnetic field.}
\end{figure}

In our calculations we mainly used vacuum-superconductor boundary
conditions $(\nabla-iA)\psi|_n=0$ and $\nabla \varphi|_n=0$ except
for the regions where current was injected (see insert in Fig. 1).
In those points we used the normal metal-superconductor boundary
conditions $\psi=0$ and $\nabla \varphi|_n=-j$.

We found two jumps in the current-voltage characteristics (see
Fig. 2). The first one is connected with the transition of the
sample from the supercondicting to the resistive state (but with
$|\psi| \neq 0$). The second transition is the one from the
resistive to the normal state. The hysteresis connected to the
second transition survives till very high magnetic field. The
latter is connected with the stability of the normal state
carrying current at temperatures lower than the critical
temperature (see for example Ref. \cite{Ivlev}). For such a system
it was shown that the normal state may exist (at $T<T_c$) till
very low current densities if there is no finite superconducting
nucleus. In our case we inject current through part of the
cross-section of the sample (see inset of Fig. 1) and in this way
we provide a more optimal condition for the nucleation of
superconductivity in the corners of the superconducting sample
where the current density is minimal even in normal state. We want
to stress that the second hysteresis crucially depends on the
geometrical parameters of the sample and it is not the subject of
the present paper to study that effect. We only study here the
first hysteresis which is almost sample-independent.

In Fig. 3 we present the dependencies of $J_{c1}(H)$ and
$J_{c2}(H)$ for different widths of the sample. We found that in
accordance with our predictions the current $J_{c1}$ increases at
low magnetic fields and at some $H=H^*$ it becomes equal to
$J_{c2}$. Beyond $H^*$ that type of hysteresis no longer exists in
the system and the critical current $J_c=J_{c1}=J_{c2}$ decreases
with increasing magnetic field. Only for relatively narrow samples
with $W\sim \xi$ there is a second peak in the $J_c(H)$
dependence. The reason for that peak is probably connected with
strong nonlinear effects. Indeed, the current density is related
to the momentum of the superconducting condensate ($p=\nabla
\phi-A$) as $j=p(1-p^2)$ in accordance to the Ginzburg-Landau
relation. The second peak occurs approximately at a field $H_s$
when the first vortices penetrate the sample in absence of
injected current. At $H=H_s$ the value of p on the edges is close
to unity for samples with $W\sim \xi$. It means that the term
$(1-p^2)$ may be very important in that range of fields. In wider
stripes the above effect is negligible and there is no second peak
in the $J_{c}(H)$ curve.

For magnetic fields larger than $H_s$, vortices will penetrate the
superconductor and occupy the central part of the sample
\cite{Kupr,Benkr}. If one injects current into the sample the
vortex-filled region starts to move to one side of the sample and
when the vortex dome touches the sample boundary the vortices will
leave the sample \cite{Kupr,Benkr}. This value of the current is
the critical one. It is essential that the distribution of the
current density in the vortex dome is practically constant in the
resistive state (in narrow samples with $W<\Lambda=\lambda^2/d$)
even in the absence of bulk pinning \cite{Maksimova}. At fields
$H\gg H_s$ the vortex dome occupies almost the whole sample and it
means that at $J\simeq J_{c1}(0)=j_{c1}W$ the conditions for the
occurrence of phase slip lines (or vortex channels) will be
fulfilled and it will lead to the appearance of a stair structure
in the I-V characteristics and to a sharp increase of the voltage
at $J>J_{c1}(0)$ \cite{Lempitsky}. It also follows that at
relatively large H this current should be field-independent. The
lower boundary of the field-independent region is defined by the
condition that the current density distribution should be almost
uniform ($H_{lower}\simeq H_s$). The upper boundary $H_{upper}$
depends on how large is magnetic field and how strong the order
parameter is suppressed in vortex filled region because according
to Refs. \cite{Kadin,Michotte} both these factors affect the value
of $j_{c1}$ (it leads to a decrease of it).

We found that the voltage sharply increases at $J>J_c$ when vortex
slip lines or vortex channels appear in the sample (compare Fig. 2
and Fig. 4 at H=0.3 and H=0.6). In contradiction with the above
prediction the current at which the vortex channelling occurs
decreases with increasing H (see Fig. 2). The most probable reason
is that the current density strongly varies over the width of the
sample for $H \gg H_s$ in case $W=9 \xi$ because even close to
$H_{c2}$ there will be only three rows of vortices (see Fig. 4).

Our results coincide qualitatively with recent experimental
results on the I-V characteristics of carbon nanotube (compare
Fig. 13 in Ref. \cite{Kasumov} with Fig. 3 in our paper) measured
at different magnetic fields. In that work a non-monotonous
dependence of $J_{c1}$ on H was observed at low magnetic fields.
In Ref. \cite{Alad} a stair structure in I-V characteristics of
Mo/Si multi-layer stripes was found which appears practically at
the same value of the injected current, as predicted in this
paper, in wide field region.

In conclusion, we found that an external magnetic field strongly
affects the resistive state of mesoscopic wires and stripes. It
leads to a shrinking of the hysteresis in the current-voltage
characteristic at relatively high values of H. If the magnetic
field is perpendicular to the current direction the dependence
$J_{c1}$(H) is a non-monotonous function of H at low magnetic
field. At high magnetic fields $J_{c1}$ and $J_{c2}$ monotonically
decreases with increasing H for both perpendicular and parallel
\cite{Michotte} orientation of the applied magnetic field.

This work was supported by IUAP, GOA (University of Antwerp), ESF
on "Vortex matter", the Flemish Science Foundation (FWO-Vl). One
of us (D.V.) is supported by DWTC to promote S $\&$ T
collaboration between Central and Eastern Europe.

\begin{figure}[hbtp]
\includegraphics[width=0.4\textwidth]{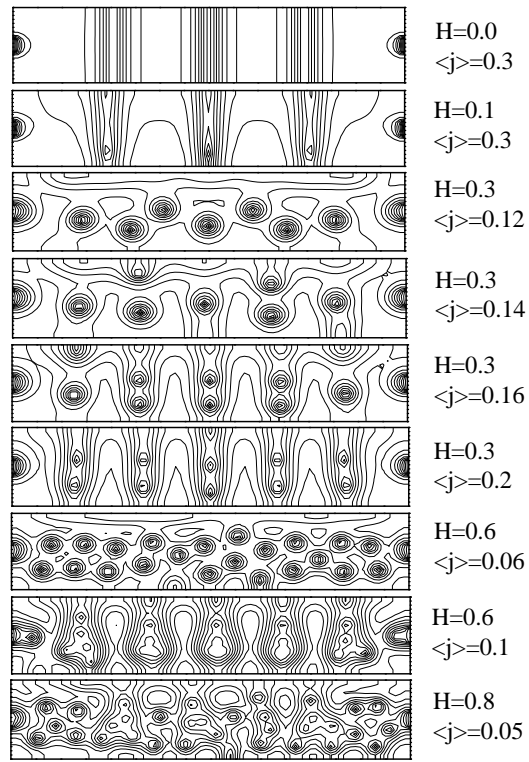}
\caption{Contour plot of the order parameter in the resistive
state of a superconducting strip at different values of the
magnetic field and applied current. The width of the sample is $9
\xi$, the length is 40 $\xi$, and we used the parameter
$\gamma=10$, $\langle j \rangle=J/(j_0W)$.}
\end{figure}


\begin{thebibliography}{20}
\bibitem{self1} Everywhere in the paper we define the current by
$J=I/d=\int_{0}^{W}j(x)dx$.
\end{thebibliography}
\end{document}